\documentclass[twocolumn,english,prl]{revtex4-1}
\usepackage[T1]{fontenc}
\usepackage[latin9]{inputenc}
\usepackage{babel}
\usepackage{amsmath}
\usepackage{amssymb}
\usepackage{graphicx}
\usepackage{wasysym}
\usepackage[unicode=true,pdfusetitle,
 bookmarks=true,bookmarksnumbered=false,bookmarksopen=false,
 breaklinks=false,pdfborder={0 0 1},backref=false,colorlinks=false]
 {hyperref}
\begin{document}
\title{Designing adiabatic time-evolution from high frequency bichromatic
sources}
\author{Álvaro Gómez-León}
\author{Gloria Platero}
\affiliation{Instituto de Ciencia de Materiales de Madrid (CSIC), E-28049 Madrid,
Spain}
\date{\today}
\begin{abstract}
We investigate the quantum dynamics of a two-level system driven by
a bichromatic field, using a non-perturbative analysis. We make special
emphasis in the case of two large frequencies, where the Magnus expansion
can fail, and in the case of a large and a small frequency, where
resonances can dominate. In the first case, we show that two large
frequencies can be combined to produce an effective adiabatic evolution.
In the second case, we show that high frequency terms (which naturally
arise as corrections to the adiabatic evolution obtained in the first
case) can be used to produce a highly tunable adiabatic evolution
over the whole Bloch sphere, controlled by multi-photon resonances.
\end{abstract}
\maketitle

\paragraph{Introduction:}

Perturbing a system out-of-equilibrium is at the heart of physics,
as it allows to extract information about its properties by just measuring
the response to the perturbation. Besides small perturbations, one
can also produce non-linear effects of high complexity, and steady
states with novel properties such as Floquet topological insulators,
skyrmions or time-crystals\citep{Floquet-TI,FitoPRL,Time-crystals,Skyrmions,Dynamical-Quantum-transitions,Engelhardt}.
Applying periodic perturbations has shown to be a versatile tool to
manipulate physical systems. For instance, they allow to control spin
qubits in quantum dots\citep{Petta2005,DynamicalSpinLocking,Transport-blocking,Rafa-dots,Jordi},
or to induce new electronic, dynamical and topological properties\citep{AlviPRL,Rudner-anomalous,Pierre1,Pierre2,Monica-pwave}.
These works typically consider monochromatic driving, although bichromatic
fields have been used in a few occasions\citep{Kholer,non-abelian,Sigmund-bichromatic,Topological-Freq-conv},
showing that their potential has not been fully explored.

The periodically driven two-level systems is one of the fundamental
models in quantum mechanics. Its physical realization has been successfully
implemented in quantum dots\citep{Rabi1,Rabi2,Rabi3,Koppens}, complex
molecules\citep{Morello,Single-ion-magnet}, superconducting devices\citep{Squid},
and many other systems\citep{Platero&Aguado}. Its universality relies
on the fact that many quantum mechanical systems, when truncated to
their low lying states by lowering the temperature, can reduce to
the dynamics between the ground state and the first excited state.

The Hamiltonian describing the unperturbed two-level system usually
displays a splitting $\Delta_{z}$. Then, one chooses this direction
as the quantization axis, and performs transitions between the ground
state and the excited state to probe the system, which is the guiding
principle in techniques such as nuclear and electron spin resonance.
The Hamiltonian describing the model can be written as $H\left(t\right)=H_{0}+V\left(t\right)$,
with
\begin{eqnarray}
H_{0} & = & \frac{\Delta_{z}}{2}\sigma_{z}\label{eq:general-eq}\\
V\left(t\right) & = & \sum_{i}\frac{V_{i}}{2}f_{i}\left(t,\omega_{i},\phi_{i}\right)\sigma_{x}\label{eq:general-eq2}
\end{eqnarray}
and where $V_{i}$, $\omega_{i}$ and $\phi_{i}$ correspond to the
different amplitudes, frequencies and phases of the external source,
respectively. The dynamics of this simple Hamiltonian can be complicated,
even in the monochromatic case, as the three energy scales involved
($\Delta_{z},V_{1}$ and $\omega_{1}$) can lead to very different
behavior. The standard perturbative analysis in $V_{1}\ll\Delta_{z},\omega_{1}$
explains the linear response regime. This is commonly used to probe
the system and obtain information about its physical properties \citep{Kubo}.
On the other hand, one can consider the (high frequency) strongly
driven regime $V_{1}>\omega_{1}$ and $\omega_{1}\gg\Delta_{z}$,
which produces the spectral changes typically found in Magnus expansions\citep{Magnus-BLANES},
and can be used to dynamically tune the properties of the system \citep{AlviPRL,FitoPRL,Dynamical-Quantum-transitions,BeaPRL,AdamePRB}.
Finally, the resonant behavior corresponds to the case $\omega_{1}\simeq\Delta_{z}$,
which produces a transfer of spectral weight from the ground state
to the excited state. This is used for state preparation in many experiments
\citep{PopulationInversion} or to induce single qubit gates\citep{Koppens}.
On top of that, environmental degrees of freedom in experiments can
also couple to the external field, producing an undesired large signal
if they are resonant.
\begin{figure}
\includegraphics[scale=0.4]{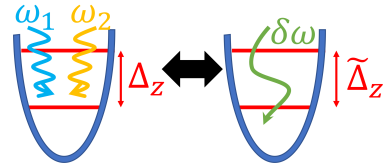}

\caption{\label{fig:Schematic}Schematic representation where a system driven
by a high frequency bichromatic field behaves as adiabatically driven
by a single frequency.}

\end{figure}

In this work we study a two-level system driven by a bichromatic field
and discuss the advantages over the monochromatic case. We analyze
the limitations of high frequency expansions, when more than one frequency
is present, and introduce a description which captures the full dynamics,
including the fine details of the micro-motion. This is in contrast
with expansions where just the stroboscopic evolution is considered,
missing the dynamics within a period, which can be crucial for a topological
analysis\citep{Nathan_Rudner}.

We find that when the two frequencies are large, but close to each
other, the field amplitudes can be controlled to produce an effective
adiabatic evolution. We demonstrate that the adiabatic behavior is
robust to noise, and highly tunable. Furthermore, we show that non-adiabatic
corrections to the resulting effective Hamiltonian can be beneficial,
and used to engineer adiabatic rotations on the Bloch sphere. We also
demonstrate that the long-time dynamics is controlled by multi-photon
resonances. In addition, our analysis can be extended to the analysis
of a multi-level system under multi-chromatic driving in a straightforward
manner.

\paragraph{Motivation:}

As an illustrative example to understand the breakdown of high frequency
expansions in a multi-frequency case, let us first consider a rather
simple trigonometric property of the function $g\left(t\right)=\cos\left(\omega t\right)$.
It can generally be written as:
\begin{equation}
g\left(t\right)=\cos\left(\omega_{1}t\right)\cos\left(\omega_{2}t\right)+\sin\left(\omega_{1}t\right)\sin\left(\omega_{2}t\right),
\end{equation}
being $\omega=\omega_{1}-\omega_{2}$ and their sum arbitrary. If
$\omega_{i}\gg\omega$, one can interpret the slowly evolving oscillatory
function $g\left(t\right)$ as coming from the difference of two large
frequencies, whose difference is very small. What in this case makes
possible to exactly map the two high frequencies to an adiabatic evolution,
is the specific relation between their Fourier components, where only
the crossed terms $g_{\pm1,\mp1}$ of the two-dimensional Fourier
expansion $g_{n_{1},n_{2}}=\int_{0}^{2\pi}\frac{d\theta_{1}}{2\pi}\int_{0}^{2\pi}\frac{d\theta_{2}}{2\pi}e^{-i\left(n_{1}\theta_{1}+n_{2}\theta_{2}\right)}g_{\vec{\theta}}$
contribute, being $g_{\vec{\theta}}$ the function $g\left(t\right)$
re-parametrized according to $\theta_{i}=\omega_{i}t$ \citep{GRIFONI1998229}.
This illustrates a specific case where a system driven by two initially
large frequencies, will not give a converging result using a high
frequency expansion. The reason is that a system with such a driving
term (with $\omega$ smaller than all the characteristic energies
of the model), obviously requires an adiabatic analysis \citep{many-mode-floquet}
due to its slow time evolution. Furthermore, it provides some intuition
about the requirements to engineer a specific dynamical behavior in
a quantum system, by studying its Fourier decomposition.

\paragraph{Bichromatic two-level system:}

Let us now move to the problem at hand. We choose a simple harmonic
protocol for the external drive $f_{i}\left(t,\omega_{i},\phi_{i}\right)=\cos\left(\omega_{i}t+\phi_{i}\right)$,
although other choices are possible ($i$ labels each different component
of the drive). One could also choose each term in Eq.\ref{eq:general-eq2}
coupled to a different, non-commuting, degree of freedom (e.g., to
a $\sigma_{y}$ component). However, this is not necessary for the
present analysis and will be discussed below. We consider the specific
case of bichromatic drive, where the sum in Eq.\ref{eq:general-eq2}
is restricted to two terms only. Each term is characterized by a different
amplitude, frequency and phase. If we parametrize $\omega_{i}t\rightarrow\theta_{i}$
to obtain a two-dimensional Fourier representation \citep{GRIFONI1998229},
the only non-vanishing Fourier components are the terms $H_{0,\pm1}$
and $H_{\pm1,0}$, which characterize the time-evolution operator
for small field amplitudes. This indicates that \textit{for weak amplitudes,
the behavior is dominated by oscillations with frequencies $\omega_{1,2}$
only}.

If instead we perform a non-perturbative analysis of $H\left(t\right)$,
by applying the transformation $\mathcal{U}\left(t\right)=\exp\left\{ -i\int V\left(t\right)dt\right\} $
we find the following transformed multi-chromatic time-dependent Hamiltonian:
\begin{eqnarray}
\tilde{H}\left(t\right) & = & \frac{\Delta_{z}}{4}\left(\sigma_{z}-i\sigma_{y}\right)\prod_{i}e^{iF_{i}\left(t,\omega_{i},\phi_{i}\right)}\label{eq:IP-Ham}\\
 &  & +\frac{\Delta_{z}}{4}\left(\sigma_{z}+i\sigma_{y}\right)\prod_{i}e^{-iF_{i}\left(t,\omega_{i},\phi_{i}\right)}\nonumber 
\end{eqnarray}
where we have defined $F_{i}\left(t,\omega_{i},\phi_{i}\right)=\int V_{i}f_{i}\left(t,\omega_{i},\phi_{i}\right)dt$,
being this result still valid for an arbitrary number of drive components.
The advantage of Eq.\ref{eq:IP-Ham} relies on the fact that, from
a Jacobi-Anger expansion\citep{BesselFunctions}, one finds non-perturbative
expressions in $V_{i}$. Notice that this type of Hamiltonian is directly
obtained in electronic systems via the Peierls substitution\citep{AlviPRL},
indicating that our results will be valid for seemingly different
systems, connected by unitary transformations. However in the latter
case, the transformation to the interaction picture is not required.

In the monochromatic case, the phase $\phi_{1}$ in Eq.\ref{eq:IP-Ham}
does not affect the spectrum, because it is a gauge degree of freedom
that sets the origin of the time-evolution, but it affects the dynamics.
However, in the multi-chromatic case the phase differences are relevant
and both, the spectrum and the dynamics are affected. This provides
an extra degree of freedom in Floquet engineering, absent in the monochromatic
case, although we will not make use of it in the present work.

The two-dimensional Fourier decomposition of Eq.\ref{eq:IP-Ham} (we
parametrize $\theta_{i}=\omega_{i}t$):
\begin{equation}
\tilde{H}_{n_{1},n_{2}}=\int_{0}^{2\pi}\frac{d\theta_{1}}{2\pi}\int_{0}^{2\pi}\frac{d\theta_{2}}{2\pi}\tilde{H}\left(\theta_{1},\theta_{2}\right)e^{-i\left(n_{1}\theta_{1}+n_{2}\theta_{2}\right)}
\end{equation}
leads to the following expression for the Fourier components of the
Hamiltonian:
\begin{eqnarray}
\tilde{H}_{n_{1},n_{2}} & = & \frac{\Delta_{z}}{4}e^{i\left(n_{1}\phi_{1}+n_{2}\phi_{2}\right)}J_{n_{1}}\left(\alpha_{1}\right)J_{n_{2}}\left(\alpha_{2}\right)\left(\sigma_{z}-i\sigma_{y}\right)\label{eq:Fourier-H}\\
 &  & +\frac{\Delta_{z}}{4}e^{i\left(n_{1}\phi_{1}+n_{2}\phi_{2}\right)}J_{-n_{1}}\left(\alpha_{1}\right)J_{-n_{2}}\left(\alpha_{2}\right)\left(\sigma_{z}+i\sigma_{y}\right)\nonumber 
\end{eqnarray}
where $\alpha_{i}=V_{i}/\omega_{i}$ and we have used the Jacobi-Anger
expansion $e^{iz\sin\left(\theta\right)}=\sum_{n=-\infty}^{\infty}J_{n}\left(z\right)e^{in\theta}$.
Eq.\ref{eq:Fourier-H} can be further simplified if the two frequencies
are commensurate \citep{BesselFunctions,Pierre1}, however we will
consider the general case.

\begin{figure}
\includegraphics[scale=0.4]{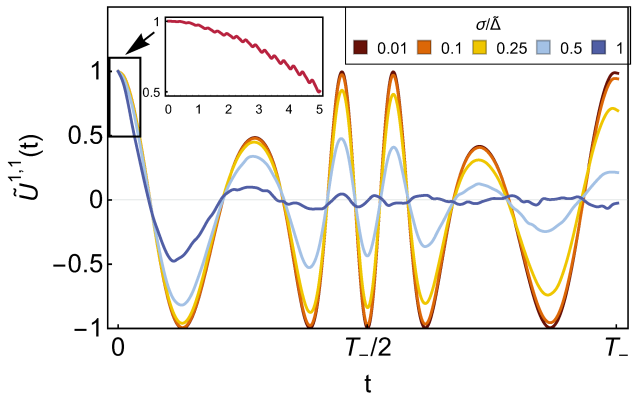}

\caption{\label{fig:Exact-Dynamics1-1}Exact dynamics during an adiabatic period
$T_{-}=2\pi/\omega_{-}$ for the real part of the upper diagonal element
of the time-evolution operator $\tilde{U}\left(t\right)$. The presence
of white noise following a normal distribution with a standard deviation
$\sigma$ has been considered. The colors indicate different noise
strength $\sigma$, in units of the dominant energy scale $\tilde{\Delta}=\Delta_{z}J_{1}^{2}\left(\alpha\right)$.
Parameters: $\omega_{1}/\Delta_{z}=10$, $\omega_{2}/\Delta_{z}=10.05$,
$\alpha\simeq1.8$ (first maximum of \textbf{$J_{1}\left(\alpha\right)$})
and $\phi_{i}=0$. The inset shows the short time dynamics, where
the fast oscillations from the fast applied drive (with period $T_{1,2}\simeq2\pi/\omega_{1,2}$)
are more evident. All the other components of the evolution operator
also evolve adiabatically.}
\end{figure}

Notice that Eq.\ref{eq:Fourier-H} contains an infinite number of
Fourier components, and in contrast with the unrealistic, but pedagogical
case of $g\left(t\right)$ above, the adiabatic behavior will only
happen if we can enhance the crossed Fourier components $\tilde{H}_{\pm1,\mp1}$.
Fortunately, due to the non-perturbative expressions in $\alpha_{i}$
this is now possible, and we choose values of $\alpha_{i}$ that maximize
$J_{1}\left(\alpha_{i}\right)$, while requiring $\omega_{-}\ll\Delta_{z}$
and $\omega_{i}\gg\Delta_{z}$ (we have defined $\omega_{\pm}=\omega_{1}\pm\omega_{2}$
and fixed $\phi_{i}=0$ for simplicity):
\begin{eqnarray}
\tilde{H}\left(t\right) & \simeq & \Delta_{z}\left[\frac{J_{0}^{2}\left(\alpha\right)}{2}-J_{1}^{2}\left(\alpha\right)\cos\left(\omega_{-}t\right)\right]\sigma_{z}\label{eq:H-full0}\\
 &  & +\Delta_{z}J_{1}^{2}\left(\alpha\right)\cos\left(\omega_{+}t\right)\sigma_{z}\nonumber \\
 &  & +\Delta_{z}J_{0}\left(\alpha\right)J_{1}\left(\alpha\right)\sum_{i=1,2}\sin\left(\omega_{i}t\right)\sigma_{y}+\ldots\nonumber 
\end{eqnarray}
Noticing that the fast oscillating term in the second line averages
to zero in this regime, and that the third line contributes with a
small amplitude correction to the time evolution, we obtain the next
leading Hamiltonian:
\begin{eqnarray}
\tilde{H}_{0}\left(t\right) & \simeq & \Delta_{z}\left[\frac{J_{0}^{2}\left(\alpha\right)}{2}-J_{1}^{2}\left(\alpha\right)\cos\left(\omega_{-}t\right)\right]\sigma_{z}\label{eq:Effective-H}
\end{eqnarray}
Actually as $\omega_{i}\gg\Delta_{z}$, all terms with a fast frequency
dependence, of the order of $\omega_{i}$ or larger, tend to zero
as $\omega_{i}\rightarrow\infty$. To demonstrate this we have numerically
calculated the time evolution operator $\tilde{U}\left(t\right)$
in the interaction picture. Fig.\ref{fig:Exact-Dynamics1-1} shows
the real part of the component $\tilde{U}^{1,1}\left(t\right)$ over
time, which is related with the occupation probability of the excited
state. Different colors indicate the time-evolution averaged over
a hundred realizations of white noise, following a normal distribution
with standard deviation $\sigma$.
 For weak noise, the plot displays adiabatic behavior (notice that
$T_{-}\sim125$ while $T_{1,2}\sim0.6$ is the period of the driving
fields) with small and fast amplitude oscillations around the mean
value, coming from the high frequency corrections (see Fig.\ref{fig:Exact-Dynamics1-1},
inset). This is one of the main results of this work: \textit{One
can combine two high frequency drives to effectively evolve the system
adiabatically}. Actually, Fig.\ref{fig:Exact-Dynamics1-1} contains
a combination of two different periods, which can be externally controlled:
harmonic oscillations due to the static part of the effective Hamiltonian
(first term in Eq.\ref{eq:Effective-H}), and a slow frequency modulation
due to the adiabatic term (second term in Eq.\ref{eq:Effective-H}).
Reducing the difference $\omega_{-}=\omega_{1}-\omega_{2}$ makes
the static part dominate at short times, while its increase makes
the non-linear, adiabatic term to take over (these two cases are explicitly
shown in the Appendix \ref{sec:Effective-Hamiltonian}, Fig.\ref{fig:Exact-Dynamics1}
with an extended discussion about the role of each term in the effective
time-dependent Hamiltonian).

In addition, the robustness of our prediction is illustrated with
the changes of the time evolution as the noise increases. Fig.\ref{fig:Exact-Dynamics1-1}
shows that the original behavior persists for weak values of noise,
until $\sigma$ becomes of the order of the dominant energy scale,
where the oscillations are strongly damped. This indicates that experiments
would have time to perform a few adiabatic cycles before the effect
of noise takes over.

\paragraph{Corrections to the effective adiabatic Hamiltonian:}

We have shown that it is possible to drive a system with two different
frequencies and produce effective adiabatic behavior. The dominant
part of the Hamiltonian, shown in Eq.\ref{eq:Effective-H}, characterizes
the slow evolution, but corrections due to oscillatory terms with
higher frequency are also present and they can produce transitions
between the adiabatic eigenstates (specially relevant are the terms
proportional to $\sigma_{y}$ in Eq.\ref{eq:H-full0}, because they
do not commute with the leading order Hamiltonian). To understand
their effect we now introduce a general formalism to study Hamiltonians
with two different frequencies, where each drive couples to a different
degree of freedom. The general form of the Hamiltonian can be compactly
written as:
\[
\tilde{H}\left(t\right)=\tilde{H}_{0}\left(t\right)+\epsilon\tilde{H}_{1}\left(t\right),
\]
where
\begin{align}
\tilde{H_{0}}\left(t\right) & =\frac{\tilde{\Delta}\left(t\right)}{2}\sigma_{z},\\
\tilde{H}_{1}\left(t\right) & =\frac{\tilde{V}\left(t\right)}{2}\sigma_{y},
\end{align}
the functions $\tilde{\Delta}\left(t\right)$ and $\tilde{V}\left(t\right)$
are, for the moment, general harmonic functions, and $\epsilon$ is
introduced to organize the perturbative series (we take $\epsilon\rightarrow1$
at the end of the calculations). To study the non-perturbative dynamics
we will calculate the time-evolution operator using multiple-scales
analysis. This method deals with the fastest time-scales first, and
then each correction characterizes new processes taking over at longer
times. Importantly, this method includes a renormalization procedure
for the secular terms \citep{Multiple-scales-book,MSA-quantum-optics,Dynamical-Spin-bath},
making the solutions valid at high/low frequency, as well as near
a resonance. This approach only neglects processes that take over
at longer time-scales than the order of the expansion in $\epsilon$.

In the first step we parameterize the time evolution operator $U\left(t\right)\rightarrow U\left(\vec{\tau}\right)$
in terms of a set of time-scales $\tau_{n}=\epsilon^{n}t$, and expand
in powers of $\epsilon$ the equation of motion for the time-evolution
operator, with $U\left(\vec{\tau}\right)=\sum_{n}\epsilon^{n}U_{n}\left(\vec{\tau}\right)$.
The equations to zeroth and first order in $\epsilon$ result in:
\begin{eqnarray}
i\partial_{\tau_{0}}U_{0} & = & \tilde{H}_{0}\left(\tau_{0}\right)\cdot U_{0}\label{eq:zeroth-equation}\\
i\partial_{\tau_{0}}U_{1}+i\partial_{\tau_{1}}U_{0} & = & \tilde{H}_{0}\left(\tau_{0}\right)\cdot U_{1}+\tilde{H}_{1}\left(\tau_{0}\right)\cdot U_{0}\label{eq:first-equation}
\end{eqnarray}
where we have omitted the $\vec{\tau}$ dependence in $U_{n}\left(\vec{\tau}\right)$.
The lowest order solution can be easily obtained from Eq.\ref{eq:zeroth-equation}
by direct matrix exponentiation:
\begin{equation}
U_{0}\left(\vec{\tau}\right)=e^{-i\int_{0}^{\tau_{0}}\tilde{H}_{0}\left(\tau_{0}\right)d\tau_{0}}\cdot u_{0}\left(\tau_{1}\right)\label{eq:unperturbed-sol}
\end{equation}
where the matrix $u_{0}\left(\tau_{1}\right)$ comes from the boundary
condition and will be determined later on, during the renormalization
procedure. Eq.\ref{eq:unperturbed-sol} can now be inserted in Eq.\ref{eq:first-equation}
and solved by choosing $U_{1}\left(\vec{\tau}\right)=u_{1}\left(\vec{\tau}\right)\cdot v_{1}\left(\vec{\tau}\right)$,
being $u_{1}\left(\vec{\tau}\right)=e^{-i\int_{0}^{\tau_{0}}\tilde{H}_{0}\left(\tau_{0}\right)d\tau_{0}}$
the solution to the homogeneous equation. The solution results in:
\begin{eqnarray}
U_{1}\left(\vec{\tau}\right) & = & -iu_{1}\left(\tau_{0}\right)\cdot\int_{0}^{\tau_{0}}\tilde{H}_{1}^{\prime}\left(\tau_{0}\right)d\tau_{0}\cdot u_{0}\left(\tau_{1}\right)\nonumber \\
 &  & -\tau_{0}u_{1}\left(\tau_{0}\right)\cdot\partial_{\tau_{1}}u_{0}\left(\tau_{1}\right)\label{eq:first-order-sol}
\end{eqnarray}
where we have defined
\begin{equation}
\tilde{H}_{1}^{\prime}\left(\tau_{0}\right)=u_{1}\left(\tau_{0}\right)^{-1}\cdot\tilde{H}_{1}\left(\tau_{0}\right)\cdot u_{1}\left(\tau_{0}\right)
\end{equation}
Eqs.\ref{eq:unperturbed-sol} and \ref{eq:first-order-sol} are the
formal solutions for the time evolution operator, which now need to
be particularized for the case of interest and renormalized, if needed.
For our present purpose we fix the specific form of the periodic functions
to $\tilde{\Delta}\left(t\right)=\tilde{\Delta}_{z}+\mu\cos\left(\omega t\right)$
and $\tilde{V}\left(t\right)=\beta\cos\left(\Omega t\right)$. In
this case $\tilde{\Delta}_{z}$ corresponds to the static part (first
term in Eq.\ref{eq:Effective-H}), $\mu$ to the dominant time-dependent
term (second term in Eq.\ref{eq:Effective-H}), and $\beta$ to the
dominant correction, non-commuting with Eq.\ref{eq:Effective-H} (the
value of the two frequencies $\omega$ and $\Omega$ is arbitrary
for the moment). The unperturbed solution is obtained from Eq.\ref{eq:unperturbed-sol}
and displays the non-linear phase evolution typically obtained in
time-dependent systems:
\begin{equation}
U_{0}\left(\vec{\tau}\right)=e^{-\frac{i}{2}\left[\tilde{\Delta}_{z}\tau_{0}+\frac{\mu}{\omega}\sin\left(\omega\tau_{0}\right)\right]\sigma_{z}}\cdot u_{0}\left(\tau_{1}\right)
\end{equation}
where $u_{0}\left(\tau_{1}\right)$ still needs to be determined.
Similarly, the first order solution in $\epsilon$ is obtained from
Eq.\ref{eq:first-order-sol}. We do not write here the full form of
the solution, because of its length and because it is enough to show
that is proportional to (details of the calculation in the Appendix\ref{sec:Multiple-scales-analysis}):
\begin{equation}
U_{1}\left(\vec{\tau}\right)\propto\beta\left[\left(n\omega\pm\tilde{\Delta}_{z}\right)^{2}-\Omega^{2}\right]^{-1}\label{eq:CorrectionH1}
\end{equation}
This correction to $U_{0}\left(\vec{\tau}\right)$ diverges if the
denominator in Eq.\ref{eq:CorrectionH1} vanishes. This is a common
feature of time-dependent perturbation theory, indicating the breakdown
of the solution, but these resonances can be renormalized in multiple-scales
analysis, and produce non-perturbative corrections to $U_{0}\left(\vec{\tau}\right)$.
Strictly speaking, the resonance condition can only be fulfilled for
commensurate frequencies (which differentiates this case with the
one of incommensurate frequencies), and for very specific values of
the parameters. However, if the denominator in Eq.\ref{eq:CorrectionH1}
becomes smaller than $\beta$, the perturbative series still diverges
and should be renormalized as well, making the difference between
incommensurate and commensurate frequencies (with very long total
period) merely a mathematical curiosity, for the physically relevant
time-scales of this setup. Therefore, one can relax the strict relation
between the parameters for a resonance, to just the approximate one:
$\left(n\omega\pm\tilde{\Delta}_{z}\right)^{2}-\Omega^{2}\lesssim\beta$. 

To renormalize the resonant terms, one needs to separate resonant
and off-resonant contributions. The amplitude corrections produced
by the off-resonant terms in Eq.\ref{eq:CorrectionH1} are of order
$\beta$, and can be neglected if we focus on $U_{0}\left(t\right)$
only. However, resonant corrections contribute to leading order and
need to be included. We assume that the system is in the regime $\omega\ll\tilde{\Delta}_{z}\ll\Omega$,
where $\omega$ corresponds to an adiabatic drive and $\Omega$ to
a high frequency one. This situation is analogous to the one obtained
in Eq.\ref{eq:H-full0} for the effective adiabatic Hamiltonian. In
this situation, several resonances can contribute (i.e., several values
of $n$ fulfill $\left(n\omega\pm\tilde{\Delta}_{z}\right)^{2}-\Omega^{2}\lesssim\beta$),
while in a different regime the analysis would be simpler, because
the resonances do not need to be included. Then, in the spirit of
multiple-scales analysis, we require that the secular terms produced
by the resonances are cancelled by $\partial_{\tau_{1}}u_{0}\left(\tau_{1}\right)$
in Eq.\ref{eq:first-order-sol}. This requirement leads to the following
flow equation for $u_{0}\left(\tau_{1}\right)$:
\begin{equation}
\partial_{\tau_{1}}u_{0}\left(\tau_{1}\right)=-i\frac{\beta}{4}\sum_{n_{0}}J_{n_{0}}\left(\frac{\mu}{\omega}\right)\sigma_{y}\cdot u_{0}\left(\tau_{1}\right)
\end{equation}
where $n_{0}$ corresponds to the set of resonances $\left\{ \pm n_{0}\right\} $
which fulfill the approximate resonance condition above. This equation
allows to determine $u_{0}\left(\tau_{1}\right)$, which encodes the
non-perturbative correction to $U_{0}\left(t\right)$. The lowest
order renormalized solution becomes:
\begin{equation}
U_{0}\left(t\right)\simeq e^{-\frac{i}{2}\left[\tilde{\Delta}_{z}t+\frac{\mu}{\omega}\sin\left(\omega t\right)\right]\sigma_{z}}\cdot e^{-it\frac{\beta}{4}\sum_{n_{0}}J_{n_{0}}\left(\frac{\mu}{\omega}\right)\sigma_{y}}\label{eq:Approximate-res}
\end{equation}
Notice that the smaller $\omega$ is, the larger is the set of resonances
$\pm n_{0}$ that needs to be included, increasing the contribution
from the non-perturbative correction. Furthermore, this correction
strongly depends on the ratio $\mu/\omega$. This indicates that if
the system is far from a resonance, or $\mu$ is not in the region
where $J_{n_{0}}\left(\mu/\omega\right)$ has a relevant weight, the
behavior is similar to that of the unperturbed solution (at least
to time-scales of the order of $\beta^{-2}$).

\begin{figure}
\includegraphics[scale=0.4]{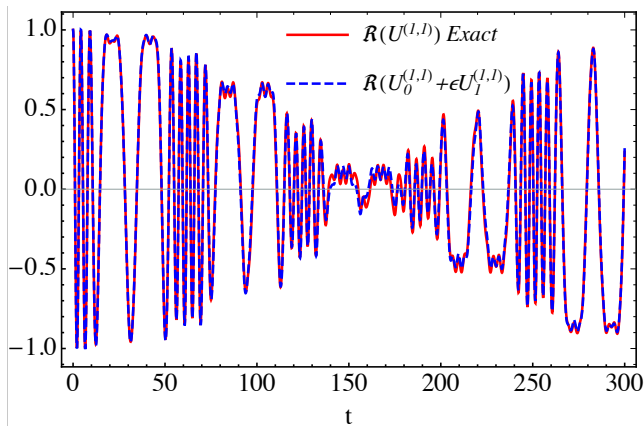}

\caption{\label{fig:Comparison2}Comparison between the exact and the approximate
dynamics to first order in $\epsilon$ for the real part of $U^{1,1}\left(t\right)$,
in the presence of resonances (Eq.\ref{eq:Approximate-res}). The
short time dynamics is well captured by the unperturbed solution,
but the slow oscillations at longer times are obtained from the renormalization
of resonances. Parameters: $\omega/\tilde{\Delta}_{z}=0.1$, $\Omega/\tilde{\Delta}_{z}=2$,
$\beta/\tilde{\Delta}_{z}=0.2$ and $\mu/\tilde{\Delta}_{z}=2$. In
this case the dominant resonance is obtained for $n=\pm10$, and its
contribution perfectly captures the slow modulation of the oscillations.}

\end{figure}

The multiple-scales analysis can be continued to higher orders in
a very systematic way, however the results presented here are quite
accurate for the range of parameters under consideration. In Fig.\ref{fig:Comparison2}
we show a comparison between the numerical and the analytical approximation
for the time-evolution operator. One can identify three different
time scales in this plot:
\begin{enumerate}
\item The shortest time-scale is given by slow harmonic oscillations coming
from the static part of the unperturbed solution $\tilde{\Delta}_{z}$
(as those oscillating between $\pm1$ in Fig.\ref{fig:Exact-Dynamics1-1}).
\item The next time-scale corresponds to the non-linear phase evolution.
Proportional to $\mu$, introduces the anharmonic oscillations happening
at intermediate times and define the adiabatic period $T_{-}$.
\item The longest time-scale is produced by the non-perturbative correction
produced by $\tilde{H}_{1}\left(t\right)$. It produces the long-time
modulation observed in Fig.\ref{fig:Comparison2} (in this case $T_{-}\sim60$
and the long-time modulation has period $\tau_{long}\sim600$, i.e.,
one order of magnitude larger).
\end{enumerate}
We also show the dynamics for the off-diagonal component of $U\left(t\right)$
in the Appendix \ref{sec:Multiple-scales-analysis}, Fig.\ref{fig:ComparisonRG2},
to confirm that the long-time behavior is controlled by the non-perturbative
correction $u_{1}\left(\tau_{1}\right)$, which produces the rotation
proportional to $\sigma_{y}$ in Eq.\ref{eq:Approximate-res}.

\paragraph{Conclusions:}

We have demonstrated that bichromatic driving provides new possibilities
to externally control quantum systems. An interesting one is that
two high frequencies $\omega_{1,2}$ can produce effective adiabatic
evolution, with frequency controlled by the difference $\omega_{1}-\omega_{2}$.
This effect requires to strongly drive the system beyond the perturbative
regime, and provides an example of the breakdown of high frequency
expansions.

The effect can be used in experiments where low frequencies are out
of reach due to equipment restrictions, or if the slow, monochromatic
drive resonantly couples to environmental (or undesired) degrees of
freedom\citep{Adiabatic-QC-NMR}. This is because the effective Hamiltonian
(Eq.\ref{eq:Fourier-H}) is a function of the coupling strength between
the drive and each degree of freedom (in this case controlled by Bessel
functions). Then, as the coupling to the environmental modes is different,
their Fourier components will be tuned at a different rate with the
field amplitude and generally suppressed, while the one of interest
is being enhanced. This would allow to reach the desired monochromatic
behavior for the degree of freedom of interest, while reducing the
undesired signal from the environment \footnote{As a final comment, notice that electronic systems coupled via the
Peierls phase to the AC source, directly lead to Hamiltonians such
as Eq.\ref{eq:IP-Ham}. This means that the transformation to the
interaction picture becomes unnecessary and our analysis directly
applies.}.

As the effective adiabatic Hamiltonian (Eq.\ref{eq:Effective-H})
generally contains high frequency corrections, we have also studied
their effect. This is equivalent to a bichromatic system with slow
and fast frequencies, coupled to different, non-commuting degrees
of freedom. In this case, we have shown that the high frequency corrections
of the effective Hamiltonian can be used to engineer controlled single-qubit
rotations. At short time-scales the adiabatic part dominates, and
one can switch between free and adiabatic evolution by adjusting the
frequency difference $\omega_{-}=\omega_{1}-\omega_{2}$. At longer
time-scales the high frequency corrections become relevant and produce
adiabatic evolution between the ground and the excited state. This
extra adiabatic evolution along a perpendicular direction is controlled
by resonances involving the slow and the fast frequencies, and its
period depends on the amplitude of both time-dependent terms ($\mu/\omega$
and $\beta$). This provides a highly tunable mechanism to implement
single-qubit gates using two off-resonant fields only.

Further applications of our results are the possibility to externally
control quantum pumping\citep{Thouless-pump,Rice-Mele-pump} in higher
dimensional systems\footnote{Work in preparation.}, or to describe
Floquet topological phases at low frequencies. This is because our
approach (multiple-scales analysis) allows a complete characterization
of the evolution operator, which is required for the topological analysis\citep{FTI-Classification}.
In qubits, it would also be interesting to study the competition between
geometric and dynamical phases, as the difference between frequencies
is tuned. This could be implemented in several experimental setups
such as quantum dots, N-V centers, single-ion magnets or superconducting
junctions.
\begin{acknowledgments}
This work was supported by the Spanish Ministry of Economy and Competitiveness
through Grant No. MAT2017-86717-P and we acknowledge support from
CSIC Research Platform PTI-001. Á. G.-L. acknowledges the Juan de
la Cierva program.
\end{acknowledgments}

\bibliographystyle{vancouver}
\bibliography{Bibliography-MSA-bichromatic}

\begin{widetext}

\appendix

\section{\label{sec:Effective-Hamiltonian}Effective adiabatic Hamiltonian
for the bichromatic two-level system}

The Hamiltonian for the two-level system driven by multi-chromatic
driving is given by:
\begin{eqnarray}
H\left(t\right) & = & H_{0}+V\left(t\right)\\
H_{0} & = & \frac{\Delta_{z}}{2}\sigma_{z}\\
V\left(t\right) & = & \sum_{i}\frac{V_{i}}{2}\cos\left(\omega_{i}t+\phi_{i}\right)\sigma_{x}
\end{eqnarray}
The non-perturbative expression in the field amplitudes is obtained
from the transformation to the interaction picture:
\begin{align}
\tilde{H}\left(t\right) & =\mathcal{U}\left(t\right){}^{\dagger}\cdot H\left(t\right)\cdot\mathcal{U}\left(t\right)-i\mathcal{U}\left(t\right){}^{\dagger}\cdot\dot{\mathcal{U}}\left(t\right)\\
\mathcal{U}\left(t\right) & =\exp\left\{ -i\int V\left(t\right)dt\right\} 
\end{align}
Simplifying the expressions, the general form can be written as:
\begin{eqnarray}
\tilde{H}\left(t\right) & = & \frac{\Delta_{z}}{2}\cos\left[\sum_{i}F_{i}\left(t,\omega_{i},\phi_{i}\right)\right]\sigma_{z}+\frac{\Delta_{z}}{2}\sin\left[\sum_{i}F_{i}\left(t,\omega_{i},\phi_{i}\right)\right]\sigma_{y}\label{eq:Interaction-picture-H}\\
 & = & \frac{\Delta_{z}}{4}\left(\sigma_{z}-i\sigma_{y}\right)\prod_{i}e^{iF_{i}\left(t,\omega_{i},\phi_{i}\right)}+\frac{\Delta_{z}}{4}\left(\sigma_{z}+i\sigma_{y}\right)\prod_{i}e^{-iF_{i}\left(t,\omega_{i},\phi_{i}\right)}
\end{eqnarray}
Considering the bichromatic case, we find that the transformed Hamiltonian
reduces to ($\alpha_{i}=V_{i}/\omega_{i}$):
\begin{eqnarray}
\tilde{H}\left(t\right) & = & \frac{\Delta_{z}}{2}\cos\left[\alpha_{1}\sin\left(\omega_{1}t+\phi_{1}\right)+\alpha_{2}\sin\left(\omega_{2}t+\phi_{2}\right)\right]\sigma_{z}\\
 &  & +\frac{\Delta_{z}}{2}\sin\left[\alpha_{1}\sin\left(\omega_{1}t+\phi_{1}\right)+\alpha_{2}\sin\left(\omega_{2}t+\phi_{2}\right)\right]\sigma_{y}\nonumber 
\end{eqnarray}
Using the Jacobi-Anger expansion in terms of Bessel functions we find
the Fourier components:
\begin{eqnarray}
\tilde{H}_{n_{1},n_{2}} & = & \int_{0}^{2\pi}\frac{d\theta_{1}}{2\pi}\int_{0}^{2\pi}\frac{d\theta_{2}}{2\pi}\tilde{H}\left(\theta_{1},\theta_{2}\right)e^{-i\left(n_{1}\theta_{1}+n_{2}\theta_{2}\right)}\\
\tilde{H}\left(\theta_{1},\theta_{2}\right) & = & \frac{\Delta_{z}}{2}\cos\left[\alpha_{1}\sin\left(\theta_{1}+\phi_{1}\right)+\alpha_{2}\sin\left(\theta_{2}+\phi_{2}\right)\right]\sigma_{z}\\
 &  & +\frac{\Delta_{z}}{2}\sin\left[\alpha_{1}\sin\left(\theta_{1}+\phi_{1}\right)+\alpha_{2}\sin\left(\theta_{2}+\phi_{2}\right)\right]\sigma_{y}\nonumber \\
\tilde{H}_{n_{1},n_{2}} & = & \frac{\Delta_{z}}{4}e^{i\left(n_{1}\phi_{1}+n_{2}\phi_{2}\right)}J_{n_{1}}\left(\alpha_{1}\right)J_{n_{2}}\left(\alpha_{2}\right)\left(\sigma_{z}-i\sigma_{y}\right)\\
 &  & +\frac{\Delta_{z}}{4}e^{i\left(n_{1}\phi_{1}+n_{2}\phi_{2}\right)}J_{-n_{1}}\left(\alpha_{1}\right)J_{-n_{2}}\left(\alpha_{2}\right)\left(\sigma_{z}+i\sigma_{y}\right)\nonumber 
\end{eqnarray}
where we have parametrized $\theta_{i}=\omega_{i}t$. To find an effective
adiabatic time evolution, we are interested in maximizing the Fourier
components which contain the frequency differences $\tilde{H}_{\pm1,\mp1}$,
while suppressing the others. If we choose the first maximum of $J_{1}\left(\alpha_{i}\right)$,
the Hamiltonian is given by ($\omega_{\pm}=\omega_{1}\pm\omega_{2}$
and we choose the two fields in phase for simplicity $\phi_{1,2}=0$):
\begin{eqnarray}
\tilde{H}\left(t\right) & \simeq & \frac{\Delta_{z}}{2}J_{0}^{2}\left(\alpha\right)\sigma_{z}-\Delta_{z}J_{1}^{2}\left(\alpha\right)\left[\cos\left(\omega_{-}t\right)-\cos\left(\omega_{+}t\right)\right]\sigma_{z}\nonumber \\
 &  & +\Delta_{z}J_{0}\left(\alpha\right)J_{1}\left(\alpha\right)\sum_{i=1,2}\sin\left(\omega_{i}t\right)\sigma_{y}+\ldots\label{eq:Full-H}
\end{eqnarray}
and its dominant contribution consists in:
\begin{eqnarray}
\tilde{H}_{0}\left(t\right) & \simeq & \Delta_{z}\left[\frac{J_{0}^{2}\left(\alpha\right)}{2}-J_{1}^{2}\left(\alpha\right)\cos\left(\omega_{-}t\right)\right]\sigma_{z}\label{eq:ApproximateH}
\end{eqnarray}
This is obtained by noticing that the high frequency terms approximately
average to zero.

In general, the effective Hamiltonian contains three types of corrections: 
\begin{enumerate}
\item Constant terms such as $\Delta_{z}J_{0}^{2}\left(\alpha\right)\sigma_{z}/2$.
They introduce a linear phase evolution for the states (see Fig.\ref{fig:Exact-Dynamics1},
blue). 
\item High frequency corrections which commute with $\tilde{H}\left(t\right)$,
such as $J_{1}^{2}\left(\alpha\right)\cos\left(\omega_{+}t\right)\sigma_{z}$.
They introduce fast oscillating non-linear corrections to the phase
evolution.
\item Non-commuting, time-dependent terms, such as $J_{0}\left(\alpha\right)J_{1}\left(\alpha\right)\sin\left(\omega_{i}t\right)\sigma_{y}$.
They produce transitions between the ground and the excited state,
and can lead to resonances.
\end{enumerate}
Nevertheless, if each independent frequency is large enough, the averaged
Hamiltonian in Eq.\ref{eq:ApproximateH} is a good approximation,
and one is left with the static and the frequency difference $\omega_{-}$
terms only, representing the adiabatic evolution. Fig.\ref{fig:Exact-Dynamics1}
shows an exact numerical simulation of the time-evolution operator
using the exact Hamiltonian in Eq.\ref{eq:Interaction-picture-H}.
\begin{figure}
\includegraphics[scale=0.38]{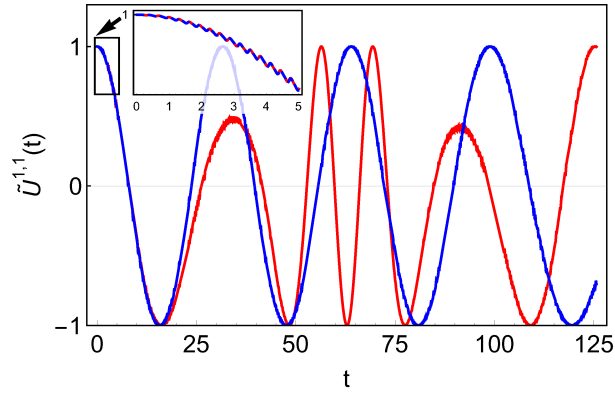}

\caption{\label{fig:Exact-Dynamics1}In red(blue), the dynamics upper diagonal
component of the time evolution operator for $\omega_{1}/\Delta_{z}=10$,
$\omega_{2}/\Delta_{z}=10.05\left(10.01\right)$, $\alpha_{i}\simeq1.8$
(this is the first maximum of \textbf{$J_{1}\left(\alpha_{i}\right)$})
and $\phi_{i}=0$. The inset shows the effect of the high frequency
corrections, as small oscillations with the frequency $\omega_{i}$
of the original drive. For $\omega_{-}=0.01$ (blue) the static part
of the effective Hamiltonian dominates at short time, displaying harmonic
oscillations with non-linearities taking over at later times (not
shown). For $\omega_{-}=0.05$ (red) the non-linear part produces
an earlier adiabatic frequency modulation.}
\end{figure}
Notice how the time evolution is adiabatic, with frequency controlled
by $\omega_{-}$, and just with small high frequency oscillations
around the mean value due to the extra harmonics. Importantly, as
the individual frequencies $\omega_{i}$ are large, compared with
the amplitudes of the extra harmonics, corrections coming from high
frequency terms are strongly suppressed.

In conclusion, if the individual frequencies $\omega_{i}$ are large,
the dynamics is controlled by just two contributions: i) the static
part, which controls the simple harmonic oscillations (blue in Fig.\ref{fig:Exact-Dynamics1}),
and ii) the non-linear adiabatic term (red in Fig.\ref{fig:Exact-Dynamics1}).
Furthermore, the ratio between these two could be independently controlled
with an extra DC field, which could enhance or suppress the static
part.

\section{\label{sec:Multiple-scales-analysis}Multiple-scales analysis for
the bichromatic case}

We consider the general time-dependent Hamiltonian:
\begin{eqnarray}
\tilde{H}\left(t\right) & = & \tilde{H}_{0}\left(t\right)+\epsilon\tilde{H}_{1}\left(t\right)\\
 & = & \frac{1}{2}\tilde{\Delta}\left(t\right)\sigma_{z}+\frac{1}{2}\epsilon\tilde{V}\left(t\right)\sigma_{y}
\end{eqnarray}
where $\epsilon$ is a dimensionless parameter which is used to organize
the perturbative series, and it is taken to one at the end of the
calculations. The time-evolution operator obeys the following equation
of motion:
\begin{equation}
i\partial_{t}U\left(t\right)=\tilde{H}\left(t\right)\cdot U\left(t\right)
\end{equation}
Now we write this equation in the usual form of multiple-scales analysis,
with parametrization $\tau_{n}=\epsilon^{n}t$. The unperturbed solution
is obtained from the unperturbed equation of motion: 
\begin{equation}
i\partial_{\tau_{0}}U_{0}\left(\vec{\tau}\right)=\tilde{H}_{0}\left(\tau_{0}\right)\cdot U_{0}\left(\vec{\tau}\right)
\end{equation}
which gives:
\begin{equation}
U_{0}\left(\vec{\tau}\right)=e^{-i\int_{0}^{\tau_{0}}\tilde{H}_{0}\left(\tau_{0}\right)d\tau_{0}}\cdot u_{0}\left(\tau_{1}\right)
\end{equation}
To first order in $\epsilon$, the equation of motion is given by:
\begin{equation}
i\partial_{\tau_{0}}U_{1}\left(\vec{\tau}\right)+i\partial_{\tau_{1}}U_{0}\left(\vec{\tau}\right)=\tilde{H}_{0}\left(\tau_{0}\right)\cdot U_{1}\left(\vec{\tau}\right)+\tilde{H}_{1}\left(\tau_{0}\right)\cdot U_{0}\left(\vec{\tau}\right)
\end{equation}
It can be solved by choosing $U_{1}\left(\vec{\tau}\right)=u_{1}\left(\vec{\tau}\right)\cdot v_{1}\left(\vec{\tau}\right)$,
being $u_{1}\left(\vec{\tau}\right)=e^{-i\int_{0}^{\tau_{0}}\tilde{H}_{0}\left(\tau_{0}\right)d\tau_{0}}$
the solution to the homogeneous equation. Then, one finds that the
solution is given by:
\begin{eqnarray}
U_{1}\left(\vec{\tau}\right) & = & -iu_{1}\left(\tau_{0}\right)\cdot\int_{0}^{\tau_{0}}\tilde{H}_{1}^{\prime}\left(\tau_{0}\right)d\tau_{0}\cdot u_{0}\left(\tau_{1}\right)-\tau_{0}u_{1}\left(\tau_{0}\right)\cdot\partial_{\tau_{1}}u_{0}\left(\tau_{1}\right)\label{eq:solution-U1}\\
\tilde{H}_{1}^{\prime}\left(\tau_{0}\right) & = & e^{i\int_{0}^{\tau_{0}}\tilde{H}_{0}\left(\tau_{0}\right)d\tau_{0}}\cdot\tilde{H}_{1}\left(\tau_{0}\right)\cdot e^{-i\int_{0}^{\tau_{0}}\tilde{H}_{0}\left(\tau_{0}\right)d\tau_{0}}
\end{eqnarray}
Now we choose specific forms for $\tilde{\Delta}\left(t\right)$ and
$\tilde{V}\left(t\right)$:
\begin{eqnarray}
\tilde{\Delta}\left(t\right) & = & \tilde{\Delta}_{z}+\mu\cos\left(\omega t\right)\\
\tilde{V}\left(t\right) & = & \beta\cos\left(\Omega t\right)\label{eq:beta}
\end{eqnarray}
where $\tilde{\Delta}_{z}$ corresponds to the static part of the
effective Hamiltonian (first term in Eq.\ref{eq:ApproximateH}), $\mu$
to the dominant periodic modulation with frequency $\omega$ (second
term in Eq.\ref{eq:ApproximateH}), and $\beta$ to the transverse
oscillating correction with frequency $\Omega$. This way, the unperturbed
solution corresponds to:
\begin{equation}
U_{0}\left(\vec{\tau}\right)=e^{-\frac{i}{2}\left[\tilde{\Delta}_{z}\tau_{0}+\frac{\mu}{\omega}\sin\left(\omega\tau_{0}\right)\right]\sigma_{z}}\cdot u_{0}\left(\tau_{1}\right)
\end{equation}
and the first order solution is obtained from the rotated Hamiltonian:
\begin{eqnarray}
\tilde{H}_{1}^{\prime}\left(\tau_{0}\right) & = & -ie^{\frac{i}{2}\left[\tilde{\Delta}_{z}\tau_{0}+\frac{\mu}{\omega}\sin\left(\omega\tau_{0}\right)\right]\sigma_{z}}\cdot\tilde{H}_{1}\left(\tau_{0}\right)\cdot e^{-\frac{i}{2}\left[\tilde{\Delta}_{z}\tau_{0}+\frac{\mu}{\omega}\sin\left(\omega\tau_{0}\right)\right]\sigma_{z}}\\
 & = & \frac{i}{2}\beta\cos\left(\Omega\tau_{0}\right)\left(\begin{array}{cc}
0 & -e^{-i\left[\tilde{\Delta}_{z}\tau_{0}+\frac{\mu}{\omega}\sin\left(\omega\tau_{0}\right)\right]}\\
e^{i\left[\tilde{\Delta}_{z}\tau_{0}+\frac{\mu}{\omega}\sin\left(\omega\tau_{0}\right)\right]} & 0
\end{array}\right)
\end{eqnarray}
To calculate the correction from Eq.\ref{eq:solution-U1}, one can
write the exponentials in terms of Bessel functions:
\begin{equation}
\tilde{H}_{1}^{\prime}\left(\tau_{0}\right)=\frac{i}{2}\beta\cos\left(\Omega\tau_{0}\right)\sum_{n}e^{in\omega\tau_{0}}\left(\begin{array}{cc}
0 & -e^{-i\tilde{\Delta}_{z}\tau_{0}}J_{-n}\left(\frac{\mu}{\omega}\right)\\
e^{i\tilde{\Delta}_{z}\tau_{0}}J_{n}\left(\frac{\mu}{\omega}\right) & 0
\end{array}\right)
\end{equation}
and perform the following integrals:
\begin{equation}
\int_{0}^{\tau_{0}}e^{i\left(n\omega\pm\tilde{\Delta}_{z}\pm\Omega\right)\tau_{0}}d\tau_{0}=i\frac{1-e^{i\left(n\omega\pm\tilde{\Delta}_{z}\pm\Omega\right)\tau_{0}}}{n\omega\pm\tilde{\Delta}_{z}\pm\Omega}
\end{equation}
These integrals will produce secular terms if $\tilde{\Delta}_{z}\pm n\omega\mp\Omega=0$,
but they can be cancelled by the term $\partial_{\tau_{1}}u_{0}\left(\tau_{1}\right)$
in Eq.\ref{eq:solution-U1}, defining the flow equation. Furthermore,
even for case where the resonance condition is approximately fulfilled
$\tilde{\Delta}_{z}\pm n\omega\mp\Omega\apprle\beta$ only, the perturbative
solution would not converge, and the renormalization can be applied.
Fig.\ref{fig:Resonance-condition} graphically shows, for a specific
case, the harmonics $n$ that require renormalization ($n=10$ in
this case). Larger $\omega$ increases the distance between different
harmonics, making more difficult to find a resonance. In this case,
beyond $\omega\gtrsim5$ resonances are no longer possible, signaling
the transition to the high frequency regime\footnote{This is not in contradiction with the first part of the paper, where
two high frequencies can give rise to adiabatic evolution. In that
case the system is far from the perturbative regime.}.
\begin{figure}
\includegraphics[scale=0.4]{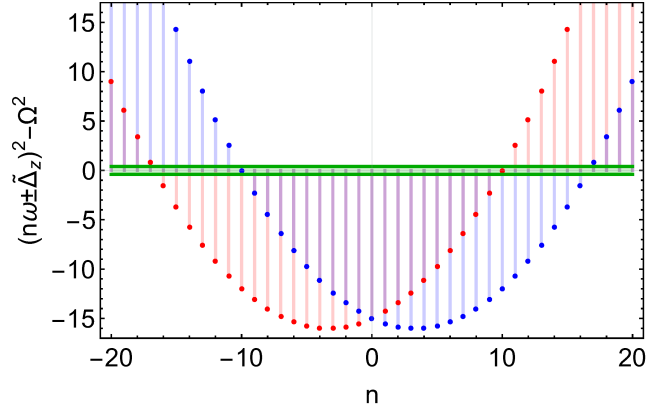}

\caption{\label{fig:Resonance-condition}Curves $\left(n\omega\pm\tilde{\Delta}_{z}\right)^{2}-\Omega^{2}$
for different values of $n$ (red and blue dots) and a region of width
$\beta$ (green, see Eq.\ref{eq:beta}). Points in the green area
fulfill the resonance condition, and require renormalization. Increasing
$\omega$ from adiabatic to diabatic values produces a transition
from several resonances to none. Parameters for the plot: $\Omega/\tilde{\Delta}_{z}=4$,
$\omega/\tilde{\Delta}_{z}=0.3$ and $\beta/\tilde{\Delta}_{z}=0.4$.}
\end{figure}
Once the subset of $n$ values which produces secular terms is identified,
the cancellation with $\partial_{\tau_{1}}u_{0}\left(\tau_{1}\right)$
in Eq.\ref{eq:solution-U1} produces the following flow equation:
\begin{eqnarray}
\partial_{\tau_{1}}u_{0}\left(\tau_{1}\right) & = & -i\frac{\beta}{4}\sum_{n_{0}}J_{n_{0}}\left(\frac{\mu}{\omega}\right)\sigma_{y}\cdot u_{0}\left(\tau_{1}\right)
\end{eqnarray}
where $n_{0}$ is the set of pairs of integers fulfilling the condition
$n_{0}\simeq\pm\frac{\Omega\pm\tilde{\Delta}_{z}}{\omega}$. Then,
the lowest order solution is given by:
\begin{equation}
U_{0}\left(t\right)\simeq e^{-\frac{i}{2}t\left[\tilde{\Delta}_{z}+\frac{\mu}{\omega}\sin\left(\omega t\right)\right]\sigma_{z}}\cdot e^{-it\frac{\beta}{4}\sum_{n_{0}}J_{n_{0}}\left(\frac{\mu}{\omega}\right)\sigma_{y}}
\end{equation}
This non-perturbative correction indicates that the oscillations of
each independent energy level are now modulated by a transition between
the ground and the excited level with frequency $\frac{\beta}{4}\sum_{n_{0}}J_{n_{0}}\left(\frac{\mu}{\omega}\right)$.
Fig.\ref{fig:ComparisonRG} compares the exact dynamics with the one
generated by $U_{0}\left(t\right)$ without/with renormalization (left/right).
The addition of non-secular contributions produces the plot in the
main text (Fig.3 main text). However, the importance of the resonances
is evident from the comparison between the left and right plot for
$U_{0}\left(t\right)$ in Fig.\ref{fig:ComparisonRG}.
\begin{figure}
\includegraphics[scale=0.4]{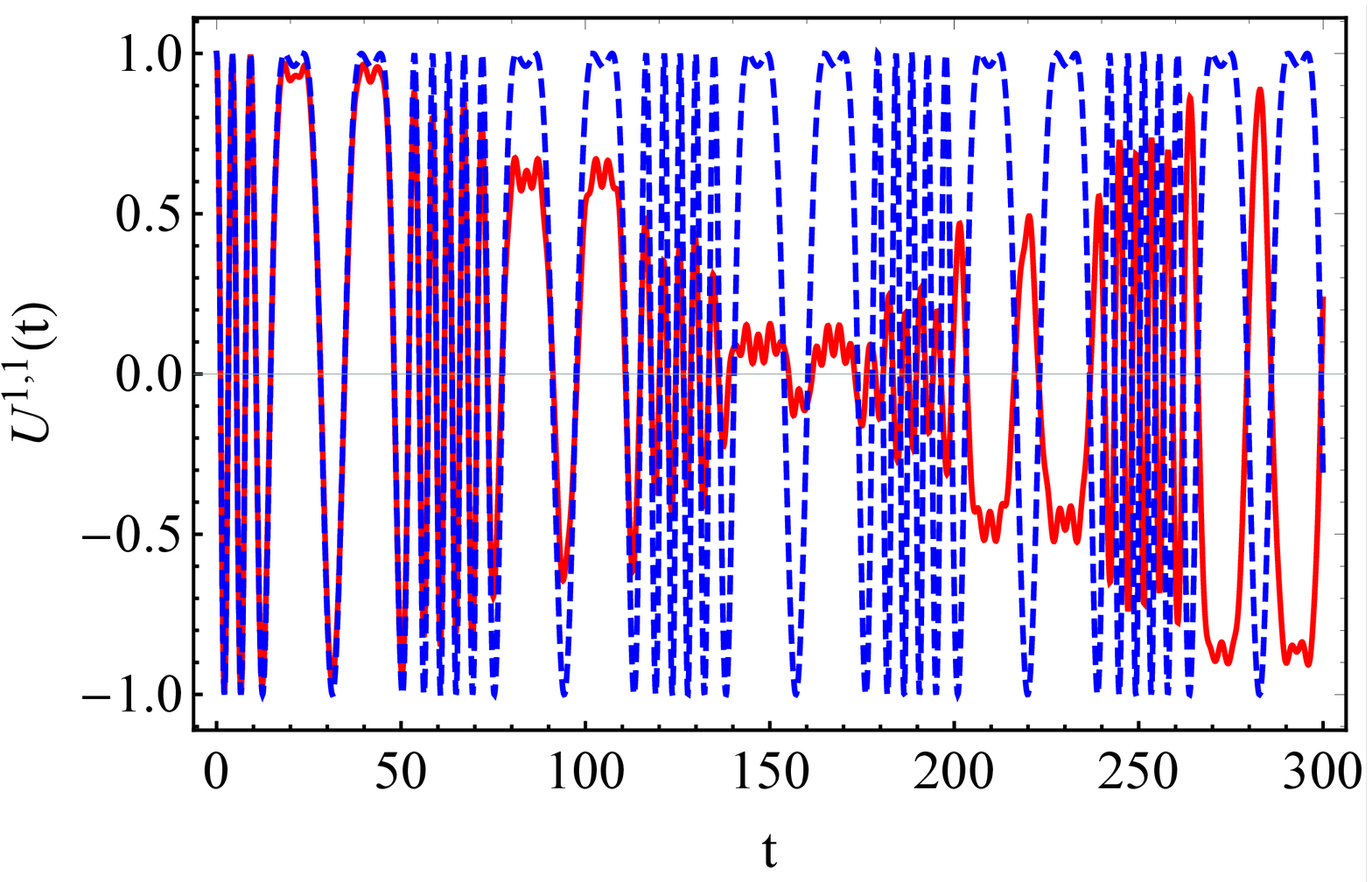}\includegraphics[scale=0.4]{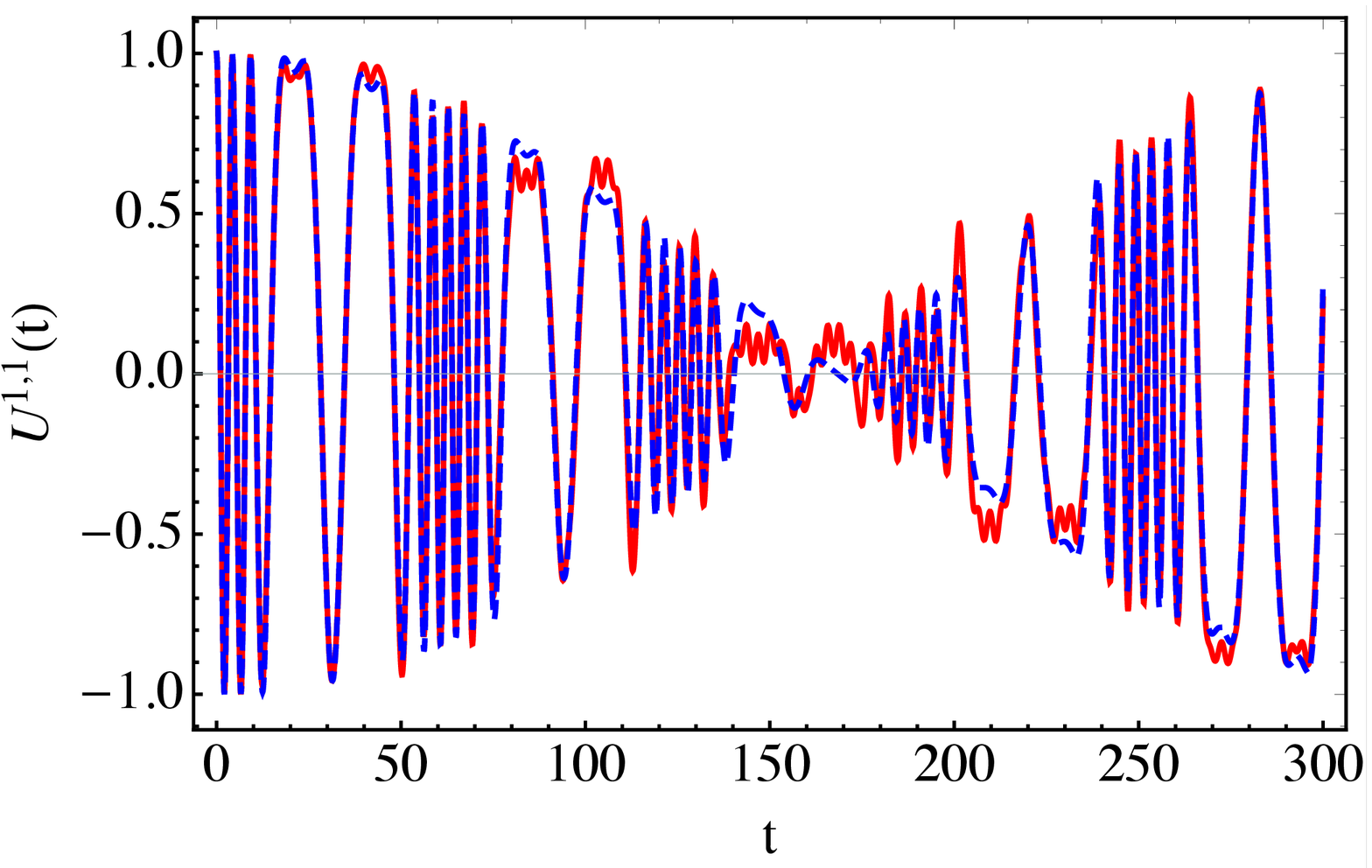}

\caption{\label{fig:ComparisonRG}Comparison between the exact dynamics (red
solid) and the one generated by $U_{0}\left(t\right)$ (blue dashed)
without/with renormalization of the resonances (left/right). We have
plotted the real part of the $U^{1,1}\left(t\right)$ component, but
the agreement is valid for all the other components as well. The addition
of small corrections of order $\beta$ to the renormalized $U_{0}\left(t\right)$
leads to an even better agreement, as shown in Fig.\ref{fig:Comparison2}
of the main text. Parameters: $\omega/\tilde{\Delta}_{z}=0.1$, $\Omega/\tilde{\Delta}_{z}=2$,
$\beta/\tilde{\Delta}_{z}=0.2$ and $\mu/\tilde{\Delta}_{z}=2$.}
\end{figure}
Finally, we plot in Fig.\ref{fig:ComparisonRG2} the exact dynamics
of $U^{1,2}\left(t\right)$, to confirm that the resonances control
the rotation proportional to $\sigma_{y}$ and that their contribution
is non-perturbative. This is confirmed by noticing that the off-diagonal
part initially vanishes, but acquires values of the order of one for
times of approximately half the period obtained from the resonances.
\begin{figure}
\includegraphics[scale=0.4]{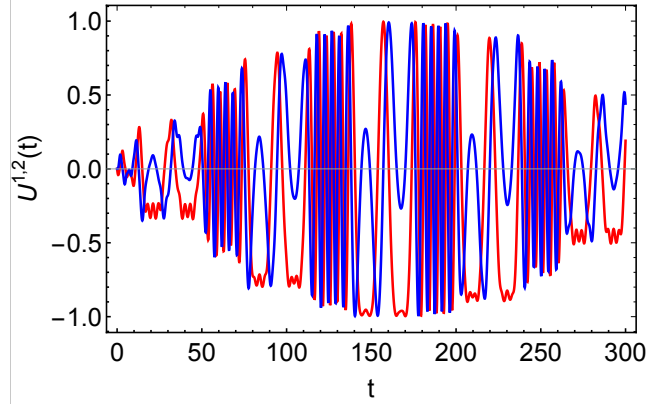}

\caption{\label{fig:ComparisonRG2}Real and imaginary parts (red and blue,
respectively) of $U^{1,2}\left(t\right)$ for the parameters $\omega/\tilde{\Delta}_{z}=0.1$,
$\Omega/\tilde{\Delta}_{z}=2$, $\beta/\tilde{\Delta}_{z}=0.2$ and
$\mu/\tilde{\Delta}_{z}=2$. One can see that the off-diagonal element
acquires non-perturbative corrections (of order one), as predicted
by the renormalization of resonances.}
\end{figure}
\end{widetext}
\end{document}